# A Thickness Dependent Enhancement of Optical Resolution in the Vicinity of an Epsilon-near-zero Slab

*Young-Rok Jang and Soo Bong Choi\**

Department of Physics, Incheon National University, Incheon 406-772, Korea

*Doo Jae Park*

Department of Physics, Hallym University, Hallymdaehaggil 1, Chuncheon 200-702, Korea

*Jisoo Kyoung‡*

Samsung Advanced Institute of Technology, Samsung Electronics Co., Suwon-si 443-803, Korea

Recent studies reports that an epsilon-near-zero (ENZ) thin slab between a specimen and a substrate contributes in enhancing the spatial resolution of the optical system. Here, we investigate the ENZ thickness dependence of the resolution enhancement. By employing the edge response function, the resolution of the optical system is directly measured when imaging a sharp edge of a metal film. We found that the optimum ENZ slab thickness was 700 nm and the achieved resolution was 11 μm at the wavelength of 8 μm. Owing to the enhanced resolution by ENZ slab, we successfully imaged the subwavelength slit arrays.

# I. INTRODUCTION

An optical resolution, how small the separation between the two objects can be discernible, is one of the most commonly used physical quantities to determine the quality of an imaging system [1]. Achieving high resolution of an optical microscope is crucial in biomolecule observation, medical diagnosis, nano-patterning, and Raman spectroscopy. In order to improve the resolution in microscopy, oil immersion lens has been widely used [2], where high-index liquid is place between the lens and the specimen to collect and deliver high-k components of the scattered light into the objective lens, resulting in a few tens of percent enhancement of the resolution. Nevertheless, oil immersion lens has a few drawbacks, including contamination of both the objective lens and the specimen. More seriously, liquid is not easy to handle, which introduces very accurate control of the horizontal balance of the microscope to guarantee the stability of the liquid.

In order to overcome those drawbacks, the solid immersion lens had been proposed [3, 4]. Usually, the solid immersion lens is made from high-index semiconductors and has spherical shape to effectively gather the high-k components of the scattered light [5]. This kind of immersion lens is used to probe the sub-surface structures of the semiconductors [6, 7] and collect the THz waves emitted or detected at the photoconductive antennas [8]. Unfortunately, however, the solid type immersion lens requires precise alignment and complicated manufacturing procedure. Recently, we have demonstrated that the optical resolution is improved when applying flat solid immersion lens or numerical aperture consist of an epsilon-near-zero (ENZ) thin slab [9], which has a vanishingly small dielectric constant at a certain frequency [10-12]. Such ENZ meta-lens has flat geometry so that no delicate optical alignment is mandatory. Also, single fabrication process without any complicated patterning is enough to make the ENZ thin film. Due to the strong directive emission between the ENZ and ordinary substrate, the high-k components are bent and gathered to the detector, resulting in a resolution enhancement. Despite the successful resolution enhancement, the dependence of the ENZ thickness has not yet been discussed. In this work, we study how the resolution is affected by the ENZ thickness and discuss about the ENZ effect in resolution enhancement in detail. Further, we will image the subwavelength structure thanks to the enhanced resolution.

## II. EXPERIMENTAL RESULTS AND DISCUSSIONS

One of the efficient ways to estimate the optical resolution of a system is to measure the edge response function (figure 1). The edge response is how the optical system responds to a sharp edge [13]. Figure 1 summarizes the measurement process of the resolution with ENZ meta-lens through the edge response. The metal half plane deposited on the ENZ film plays a role in the sharp edge. The incident plane wave will be distorted and the transmitted wave has a step-like shape along the vertical direction of the edge. Then, the distance required for the edge response to rise from 10 % to 90 % is a single parameter expressing the resolution of the optical system, which means the minimum separation to distinguish the two objects. Therefore, large distance means poor resolution. We will call this parameter the width of the edge response, . Using the edge response for measuring resolution has many advantages. First of all, the resolution can be estimated from the measured image information itself. Second, the imaging of the half plane is quite simple and the line spread function as well as modulation transfer function can easily be found from the edge response function. To measure the resolution enhancement enabled by the ENZ materials, various thicknesses of the $SiO_2$ film from 100 all the way up to 900 nm were deposited on the Si substrate, which is a natural ENZ material at 8 μm. The thickness of the metal pattern made from Au and the SiO2 film were 70 nm and 300 nm, respectively.

The edge response function was recorded by the focal plane array detector with a pixel resolution of 2.7 μm and a covering area of 170 μm by 170 μm. Since each pixel can measure total transmission spectra, it is possible to obtain the width of the edge response function at a certain wavelength (figure 2).

The value of $\delta$ as a function of incident light wavelength is plotted in figure 2 (a) with (red circles) and without (black squares) ENZ film. It is well-known that the resolution is getting worse as the wavelength increases and this is verified by the resolution measurement without ENZ (black line in figure 2 (a)). With ENZ film, the resolution apart from the ENZ position is bad compared to the Si substrate (red circles in figure 2 (a)). In contrast, $\delta$ suddenly drops around the ENZ position and reaches 12 μm at the 8-μm-wavelength. This means that the resolution can be enhanced by a factor of 1.5 only by inserting an ENZ material between the specimen and the substrate. The resolution enhancement originates from the strong directive emission phenomena at the ENZ and substrate junction [9]. To support this statement, we also measured the resolution at reflection side (figure 2 (b)). Remarkably, the edge response width as a function of wavelength with (red circles) and without (black squares) the ENZ material are almost similar. This is because the directive emission does not exist at the reflection side and thereby the high-k components of the scattered light are not able to reach the detector regardless of the existence of ENZ slab.

The resolution enhancement should depend on the thickness of the ENZ film because too thin film would be negligible losses while too thick film are not. In order to find the optimum thickness, we measure the width of the edge response by changing the ENZ film thickness (figure 3). The thickness of ENZ film was varied from 100 nm to 900 nm with 200 nm intervals and the $\delta$ was measured at 8-μm-wavelength. Roughly speaking, the resolution is getting better until the thickness of the ENZ film reaches 700 nm, but the difference is only about few percent. The minimum width of the edge response was recorded 11 μm which is not that much larger than the wavelength (8 μm). Thus, we expect that further optimization with the low loss ENZ film might allow achieving the subwavelength resolution.

Finally, we apply our meta-lens for imaging subwavelength structures. The target subwavelength structure consists of 1-μm-width slit array with the period of 50 μm (figure 4 (a)). Figure 4 (b) shows the full spectral wave imaging of the sample. Due to the subwavelength nature, the slits are hardly discernable. Since we measure the full spectrum of the transmitted light at each pixel, we can extract the intensity map at 8-μm-wavelength from this image. Remarkably, the subwavelength slits are distinguishable at the ENZ position as clearly seen in the figure 4 (c). Two image maps are drawn on the basis of the absorbance so that the metal part has red color rather than the opening part. The absorbance can be converted to transmission *T* as follows,

$$\alpha = 2 - \log_{10} T \tag{1}$$

From the cross-section plot of the figure 4 (c) (along the dotted line), we can verify the period of the slits as 50 μm (figure 4 (d)).

## III. CONCLUSIONS

In conclusion, we introduce the edge response function for measuring the resolution of the optical system. The resolution is estimated from how blurred the image of the sharp metal edge is. The rising from 10 to 90 % of the edge response is inversely proportional to the resolution of the system. Based on the edge response function we compare the resolution with and without ENZ-meta lens. The 1.5 times resolution enhancement with the ENZ film is observed at the transmission side owing to the strong directive emission between the ENZ and Si substrate. In contrast, at the reflection side, the resolution enhancement disappears because the most of the light reflects back at the surface of the sample and thereby no directive emission exists. By changing the thickness of the meta-lens, we find that the 700-nm-thick ENZ film is the optimum thickness for highest resolution enhancement. The measured minimum width of the edge response was 11 μm and further study is needed to achieve the subwavelength resolution. Based on the resolution enhancement, we successfully imaged the subwavelength slit arrays at the ENZ position. Since our scheme for the resolution enhancement is simple, we believe that our method can be directly applicable to conventional microscope. In addition, we expect that the resolution enhancement can be possible in the visible range with specially designed metamaterials [14, 15].

## ACKNOWLEDGMENT

This research was supported by Incheon National University Research Grant in 2013.

## REFERENCES

[1]   G. R. Fowles, Introduction to Modern Optics (1989).
[2]   M. Switkes and M. Rothschild, "Resolution enhancement of 157 nm lithography by liquid immersion," MOEMS 1, 225-228 (2002).
[3]   L. Novotny and B. Hecht, Principles of Nano-optics (Cambridge University Press, 2006).
[4]   S. M. Mansfield and G. S. Kino, "Solid immersion microscope," Appl. Phys. Lett. 57, 2615-2616 (1990).
[5]   Q. Wu, G. D. Feke, R. D. Grober, and L. P. Ghislain, "Realization of numerical aperture 2.0 using a gallium phosphide solid immersion lens," Appl. Phys. Lett. 75, 4064-4066 (1999).
[6]   S. B. Ippolito, B. B. Goldberg, and M. S. Ünlü, "High spatial resolution subsurface microscopy," Appl. Phys. Lett. 78, 4071-4073 (2001).
[7]   A. Yurt, A. Uyar, T. B. Cilingiroglu, B. B. Goldberg, and M. S. Ünlü, "Evanescent waves in high numerical aperture aplanatic solid immersion microscopy: Effects of forbidden light on subsurface imaging," Opt. Express 22, 7422-7433 (2014).
[8]   J. S. Kyoung, M. A. Seo, H. R. Park, K. J. Ahn, and D. S. Kim, "Far field detection of terahertz near field enhancement of sub-wavelength slits using Kirchhoff integral formalism," Opt. Commun. 283, 4907-4910 (2010).
[9]   J. Kyoung, D. J. Park, S. J. Byun, J. Lee, S. B. Choi, S. Park, and S. W. Hwang, "Epsilon-Near-Zero meta-lens for high resolution wide-field imaging," Opt. Express 22, 31875-31883 (2014).
[10]  M. Silveirinha and N. Engheta, "Tunneling of Electromagnetic Energy through Subwavelength Channels and Bends using ε -Near-Zero Materials," Phys. Rev. Lett. 97, 157403 (2006).
[11]  A. Alu, F. Bilotti, N. Engheta, and L. Vegni, "Metamaterial covers over a small aperture," IEEE Trans. Antennas. Propag. 54, 1632-1643 (2006).
[12]  A. Alù, M. G. Silveirinha, A. Salandrino, and N. Engheta, "Epsilon-near-zero metamaterials and electromagnetic sources: Tailoring the radiation phase pattern," Phys. Rev. B 75, 155410 (2007).
[13]  S. W. Smith, The Scientist & Engineer's Guide to Digital Signal Processing (California Technical Publishing, San Diego, CA 1997).
[14]  E. J. R. Vesseur, T. Coenen, H. Caglayan, N. Engheta, and A. Polman, "Experimental Verification of n=0 Structures for Visible Light," Phys. Rev. Lett. 110, 013902 (2013).
[15]  R. Maas, J. Parsons, N. Engheta, and A. Polman, "Experimental realization of an epsilon-near-zero metamaterial at visible wavelengths," Nat. Photon. 7, 907-912 (2013).

**FIGURES**

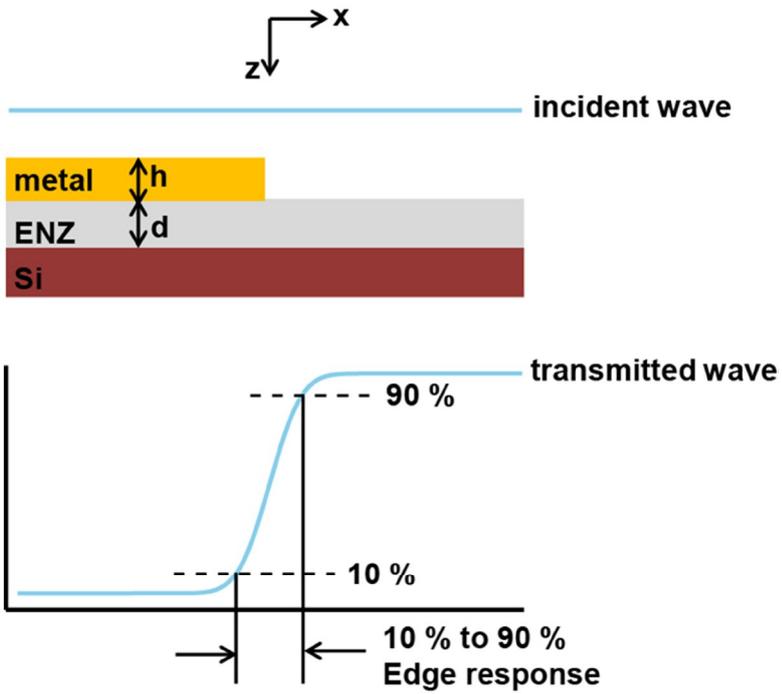

**Fig. 1.** Schematic diagram of the edge response function and its measurement.

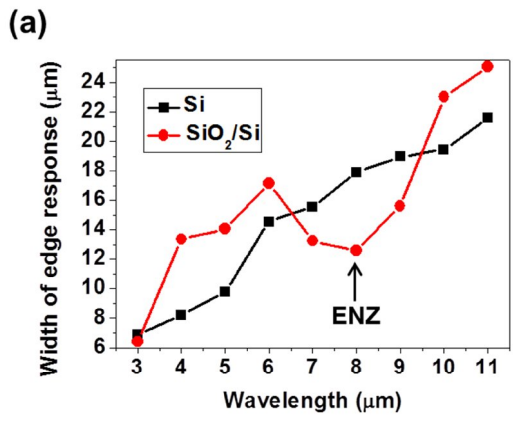 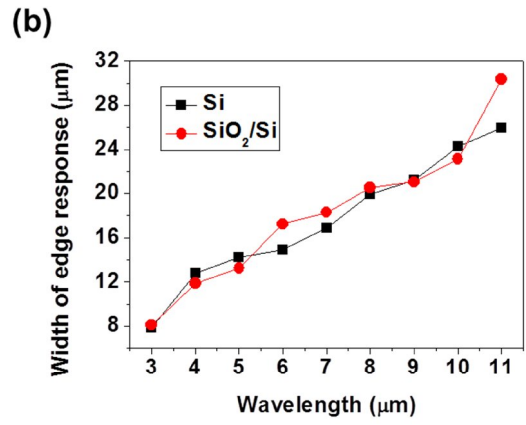

**Fig. 2.** Measured the width of the edge response with (red circles) and without (black squares) the ENZ material at (a) transmission side and (b) reflection side. Large width means poor resolution.

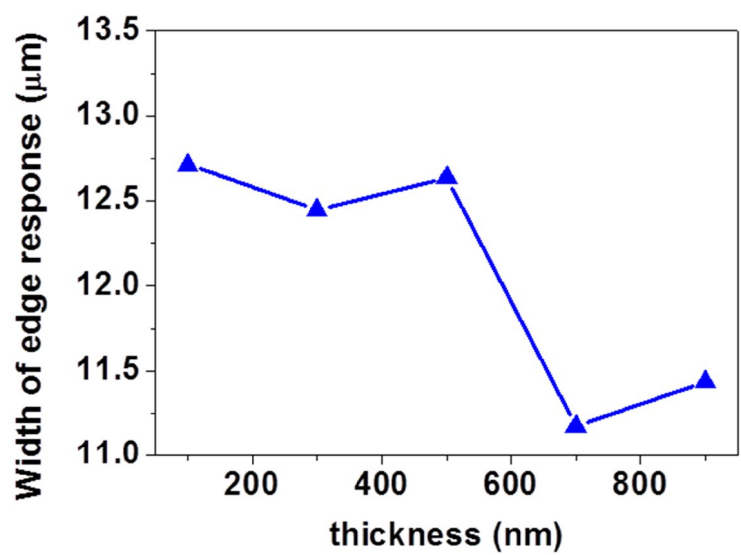

**Fig. 3.** ENZ film thickness dependence of the edge response width.

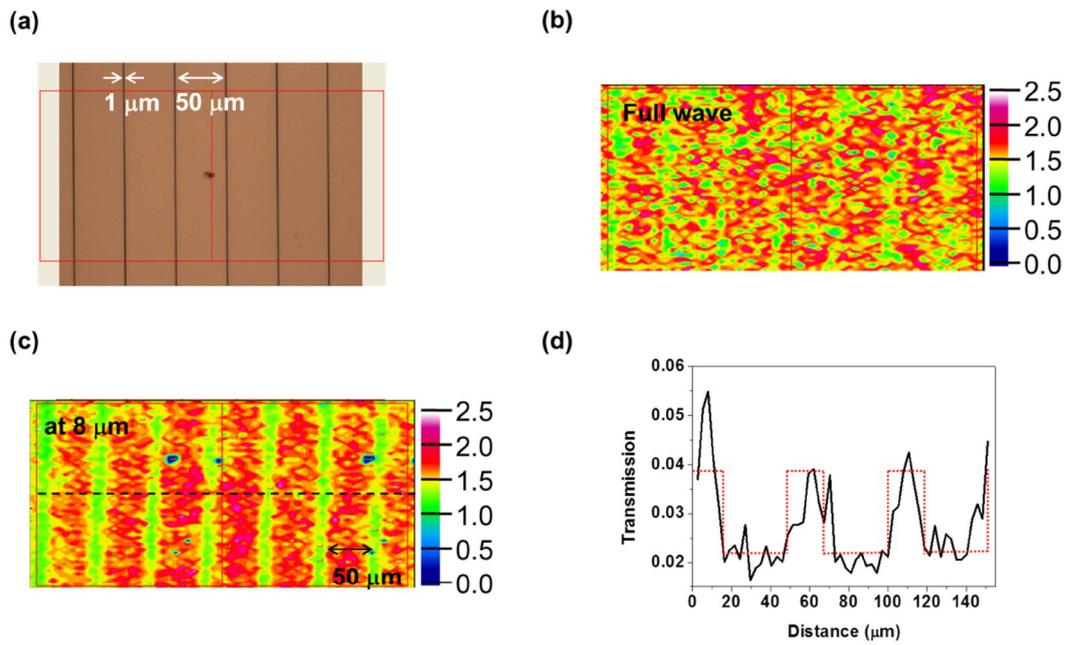

**Fig. 4.** (a) A microscope image of the subwavelength slit arrays. (b) Full spectral absorbance image. (c) absorbance image at 8- μm -wavelength. (d) Transmission plot along the dotted line in (c).